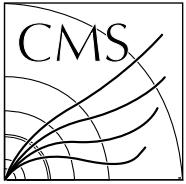

## The Compact Muon Solenoid Experiment

# CMS Note

Mailing address: CMS CERN, CH-1211 GENEVA 23, Switzerland

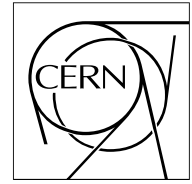

# From Design to Production Control Through the Integration of Engineering Data Management and Workflow Management Systems

J.-M. Le Goff

*CERN, Geneva, Switzerland*

G. Chevenier[1]

*Institut fur Teilchenphysik, Eidgenossische Technische Hochschule, Zurich, Switzerland*

A Bazan, T. Le Flour, S. Lieunard, S. Murray, J.-P.Vialle

*LAPP, IN2P3, Annecy-le-Vieux, France*

N. Baker, F. Estrella[1], Z. Kovacs[1] R. McClatchey[1]

*Dept. of Computing, Univ. West of England, Frenchay, Bristol BS16 1QY UK*

G. Organtini

*Universita di Roma e Sezione dell' INFN, Roma, Italy*

S. Bityukov

*IHEP, Protvino, Russia*

### Abstract

At a time when many companies are under pressure to reduce "times-to-market" the management of product information from the early stages of design through assembly to manufacture and production has become increasingly important. Similarly in the construction of high energy physics devices the collection of (often evolving) engineering data is central to the subsequent physics analysis. Traditionally in industry design engineers have employed Engineering Data Management Systems (also called *Product Data Management Systems*) to coordinate and control access to documented versions of product designs. However, these systems provide control only at the collaborative design level and are seldom used beyond design. Workflow management systems, on the other hand, are employed in industry to coordinate and support the more complex and repeatable work processes of the production environment. Commercial workflow products cannot support the highly dynamic activities found both in the design stages of product development and in rapidly evolving workflow definitions. The integration of Product Data Management with Workflow Management can provide support for product development from initial CAD/CAM collaborative design through to the support and optimisation of production workflow activities. This paper investigates this integration and proposes a philosophy for the support of product data throughout the full development and production lifecycle and demonstrates its usefulness in the construction of CMS detectors.

[1] Now at CERN

# 1. Introduction

In the process of product development, spanning design to implementation, the management of product design information is paramount. This is especially true when manufacturers are having to optimise their process engineering so that product development times and "times-to-market" are reduced. As the complexity of products increases, and these days composite products are being manufactured with hundreds of thousands of component parts, so does the requirement for the use of computer-based management products. Furthermore, distributed production of products requires that product data and documents be available across local- and wide-area networks and that there is coordinated access to the product data.

Engineering Data Management (EDM) tools have been used for some time by manufacturing companies such as Mercedes-Benz and Ford to manage the data and documents accumulated in the design of their products. These systems are normally based on commercially available EDM Systems (EDMS) such as Adra's Matrix, IBM PM or Sherpa. The features of EDM Systems include data vault and document management, product structure management and project program management [Phi96]. EDM Systems have been successfully employed to control the documents and data files emerging from the creative and collaborative stages of product design (e.g. CAD/CAM data) where product structures often tend to be hierarchical in nature and when access to documents needs to be controlled between groups of designers (such as in e.g. folder management).

The advantages of using an EDMS are well-documented elsewhere [Phi96, Ham96, PPS96]. With an EDMS the so-called 'product breakdown structure' (PBS) data is centralised, versioned and can be used for tracking design in a environment which supports collaboration on creative work. EDMSs provide a change management service which can be used by engineering applications to assess, control and minimise the impact of material, product and process changes that occur in complex manufacturing lifecycles. EDM systems comprise a set of integrated applications that improve the efficiency of people and processes involved in the design, production and assembly of system parts. However, although EDM Systems provide good support for product documents and data particularly at the early stages of design, their use in supporting the processes inherent in *product development* is somewhat limited [PM96]. Also PDM systems provide no facilities for the enactment or execution of *production* and *assembly* activities.

Workflow management systems [GHS95], on the other hand, allow managers to coordinate and schedule the activities of organisations to optimise the flow of information and operations between the resources of the organisation. Commercial workflow management systems and research products are becoming available for the storage of workflow-related information and in the capture of so-called *audit trails* or event trails of workflow operations. These systems seem to be appropriate tools for supporting the enactment of defined workflow operations. Workflow systems are, however, weak at handling the dynamic evolution of process definitions which occurs during the *design* process and can occur even during the enactment of workflow processes (eg. in so-called scientific workflow management).

Typically therefore, in manufacturing systems, design engineers use an EDMS and production managers use Production Planning Systems and/or workflow management software. Design control and production control are separated and there is little or no cross-talk between the two. This despite the fact that design changes need to be reflected quickly into the production environment to reduce development time. The provision of continuity from design to production through the capture of consistent part data is therefore a high priority. The integration of EDMS with workflow management software to facilitate consistency and continuity seems appropriate.

Up to now the integration of workflow and product data management has only been proposed for the capture of simple design information in manufacturing [Ram96]. This paper outlines how PDM and Workflow management systems can be integrated to facilitate the support of the full product development lifecycle in manufacturing from design through to operation of (versions of) the final production line. The example used in this paper draws on experience in the development of the CRISTAL scientific workflow system which is being carried out at CERN, the European Centre for Particle Physics. The CRISTAL project [LeG96] is being developed to manage the data collected from and the dynamic processes invoked in the construction and assembly of the Compact Muon Solenoid (CMS [CMS95]) experiment due to be run at CERN from 2005 onwards. The nature of the construction of CMS is heavily constrained by time and budget, is research-based (and thus apt to evolve rapidly) and is highly distributed. The overall performance of the CMS detector is highly dependent on its design and, as a consequence, any changes in design need to be permeated through from conceptual design to physical construction as quickly as possible (see figure 1 for the description of the detector product development lifecycle). It is therefore a challenging and appropriate environment to investigate EDM and workflow integration. Further detail on CRISTAL specifics can be found in [LeG96, McC97a & McC97b].

This paper firstly investigates the role of Product Breakdown Structures and the use of EDMS (or together the 'As-Designed' view of part data) in the definition of the construction of the CMS Electromagnetic Calorimeter (ECAL). Then it introduces the concept of meta-data to simplify data capture in the Assembly Breakdown Structure (ABS) or the 'As-Built' view of part data . The following section addresses issues behind the integration of workflow management with an EDMS. Next the paper shows how we can define packages to handle data formats, agents and execution specification in the CRISTAL project. The penultimate section shows how CRISTAL handles versioning of the production process and the final section covers discussion of more general use of CRISTAL in CMS.

## 2. Product Breakdown Structure PBS (Detector 'As-Designed')

The development lifecycle of a large high-energy physics detector is much like any other large-scale manufacturing activity in that it follows a design-prototype-implement cycle (see figure 1). Since the nature of detector construction is highly dependent on state-of-the-art materials and techniques it does differ from industrial production in that it is highly iterative and consequently dynamic in execution. At the outset of the development a study is carried out (on the basis of some simulations) which assesses the feasibility of detector construction. The mechanical design of the detector is similarly dictated by the choice of materials as well as physics considerations.

CAD/CAM systems (e.g. Euclide at CERN) are normally employed by mechanical engineers to specify the design of individual detector components or parts. Conceptual design tends to be a collaborative activity between groups of designers with individual designers checking-out and checking-in documents and diagrams of components under a policy of controlled configuration management [Fei91]. The database (and data vault) aspects of an EDMS lend themselves well to this creative design process (figure 1). In EDM systems Product Breakdown Structures (PBS), or as industry refers to them, "Bills Of Material" (BOM), are often strictly hierarchical in form (e.g. Car-engine-piston) and attributes can be assigned to each part or sub-part in turn in the hierarchy. Objects located in the product hierarchy can go through several stages of development so that "state" can be assigned to a part and this can be managed by the EDMS. Commercial EDMS products give adequate coverage in all these areas of functionality.

One consequence of using an EDMS as a central repository for part-related data is that there are often multiple viewpoints on part-related data in an enterprise. Designers are concerned with *types* of parts rather than final assembled parts whereas constructors must deal with *actual* produced parts. Similarly in high energy physics detector construction there are two distinct viewpoints: the 'as designed' view of the detector (embodying the Product Breakdown Structure, PBS) comprising thousands of *types* of parts defined by design engineers and the 'as built' view of the detector (or Assembly Breakdown Structure, ABS) comprising the physical detector constructed of millions of *actual* parts. The 'as-designed' view of the detector embodies the engineers design knowledge and allows the specification of tools, 3-dimensional drawings, generic module definitions etc. Whereas the 'as-built' viewpoint is needed by the physicist since each actual assembled detector component will have its own individual geometrical position and physical characteristics (such as transmission and light yield for the ECAL crystals) which are essential to provide the physics measurements from the detector.

In essence the purpose of building a PBS for industrial applications in an EDMS is to facilitate the capture of a design *hierarchy* of hundreds of parts. The PBS enables the design of individual product constituents and provides a structure for ordering and cost tracking. The PBS could be constructed from *actual* part names (e.g in the CMS ECAL: Super-Module 25) but, as the final product becomes more and more complex in structure, this naming approach leads to a parts explosion in the EDMS and data management becomes a problem. In the example of CRISTAL, the CMS detector will be constructed from millions of individual parts many of which will be identical in nature (e.g. ECAL has 36 identical Super-Modules and therefore 36 Super-Modules in the 'as-built' view). It is simply infeasible to enter, name and manage all parts individually in the EDMS.

Instead, a concept of meta-data management can be followed to alleviate this problem where *definitions* of parts (i.e. *types* such as Super-Module_Definition) are captured in the EDMS and instantiations of these definitions are (re-)used to form the PBS (see figure 2). This concept of storing types of parts (the 'as-designed' view) rather than individual copies of identical parts (the 'as-built' view) reduces the database to a manageable level without losing any information inherent in the PBS. A simple calculation reveals that, even excluding controls, monitoring and cabling information for the CMS ECAL, this meta-data approach requires the management of only around 400 definitions; without meta-data EDMS would need to manage around 250,000 ECAL parts.

# 3. Assembly Breakdown Structure ABS (Detector 'As-Built').

A product breakdown structure (PBS) approach to design has already been advocated by Bachy & Hameri [BH95] in the design of the LHC [LHC93] accelerator. According to [BH95], any EDMS system used for the engineering of large-scale one-of-a-kind facilities should hold the descriptions of the PBS, the assembly breakdown structure (ABS) and the work breakdown structure (WBS). Whereas the PBS saves information pertaining to projects, sub-projects, documents, items etc., the ABS holds data about how actual component parts (and composite parts) are assembled to form the overall final product (CMS in CRISTAL). This section of the paper describes how the 'as-designed' and 'as-built' views of part information or the PBS and ABS are related. Discussion of the WBS is deferred to the next section of this paper.

To describe how part data can be viewed from different perspectives (or viewpoints) consider the example of the CMS ECAL elements as shown in figures 3. The CMS design engineers, in their view of the part data, will specify, amongst others, types of Sub-Modules (e.g. Sub-Module Type 1), Sub-Units (Type 1Left, Type 1Right), Alveolar Structures (Type 1), Tablets, Crystals (Types 1Left and 1Right) and Capsules. They will also specify the numbers of each which form the detector (e.g. Sub-Module Type 1 is comprised of 5 Sub-Units of Type 1Left, 5 Sub-Units of Type 1Right, an Alveolar Structure of Type 1and 1 Tablet). Together this information can be displayed in the PBS graph as shown in figure 3a where data is represented as a "Bill of Materials" structure with tagged quantity attributes whose data is manageable in quantity (detector 'as-designed'). Those responsible for the production and assembly of the ECAL elements will, however, view the data from a standpoint of actual detector components (with associated physical characteristics) so will need to access an expanded version of this graph. Individual components will need to be isolated and information, such as geometry, gathered for them (e.g. Sub-Unit 1Left #2 will be slotted into a position different from Sub-Unit 1Left #3). As a consequence, the physicists view the part data in the hierarchical form of the ABS as shown in figure 3b (detector 'as-built').

Thus in implementing a meta-data management approach, the rigid tree representation of the Product Breakdown Structure is transformed into a graph representation from which an actual Assembly Breakdown Structure (ABS) can be generated. Each node in the ABS is an actual part in the detector and the node position in the tree represents its effective location in the experiment. Individual parts (of the same type) are then associated with common definitions, avoiding a parts explosion. This mechanism provides ease of data management in that there will be many fewer objects to manage, but it is still possible to derive the full tree structure for assembly from the graph representation when necessary.

Figure 4 shows an UML [UML97] object model of the product breakdown structure which underlies CRISTAL. A Part Definition is the entry point for all the information required to produce, build and characterise any part of a detector which has been registered in the system with that definition. Part Definitions are either Elementary, Composite in nature or are a Tool Definition. A Tool Definition describes the tool which is required to build the composite part. Composite parts are made up of other parts and the PartCompositionMember object reflects membership of a part in a composite.

# 4. Work Breakdown Structures - the Integration of EDMS with Workflows

Whereas the PBS and ABS describe the design of generic parts and the assembly of actual parts, the WBS holds information about the organisation of tasks (or activities) to be performed on parts and the resources required for each task. The WBS defines the activities which enable the engineers to build the production line which can be viewed as a collection of (versioned) workflows. The ABS and WBS together hold the definitions of the production line and can be mapped onto workflows (see following section).

A workflow management system (WFMS) according to [GHS95, Sch96] is a system that completely defines, manages and enacts workflows through the execution of software whose order of execution is driven by a computer representation of the workflow logic. Workflows are collections of human and machine based activities (tasks) that must be coordinated in order that groups of people can carry out their work. In CRISTAL it is therefore the workflow system that "glues" together the different organisations, operators, processes, data and centres into a single coordinated and managed production line. The Workflow Definitions are built from the assembly sequence diagrams that are provided by mechanical engineers. CRISTAL is used to support information collected during the prototyping, construction and maintenance activities of the product lifecycle (see figure 1) and to facilitate quality control and production management.

The main components of a workflow management system are a workflow application programming interface and a workflow enactment service. The workflow application programming interface allows for the specification of workflows and workflow activities (which may be composite in nature) and for the specification of activities to resources (people, machines etc.). The workflow enactment service consists of an execution interface and an execution service provided by a so-called workflow engine. The engine is the component that executes the static workflow (production) descriptions which were defined through the programming interface and provides run-time services capable of managing and executing those instances. Workflow products normally subsume both the specification and enactment of workflows. In essence workflow systems hold information about work breakdown structures (WBS) and details of execution specifications.

The CRISTAL system therefore comprises two distinct functions: one of product data management (EDMS) and one of workflow management. To achieve integration between the EDMS and workflow components of CRISTAL, the EDMS can be used to store a sets of definitions (or a design model) of both the parts and the tasks that need to be executed on the parts. In manufacturing systems the EDMS can then manage the *definitions* of the product and workflow data and the Workflow software can cater for the *instantiation, scheduling and enactment* of those definitions. The EDMS acts as the reference database both for the activation and enactment services of the production workflow system and for other systems (eg. mechanical engineering) and manages (versions of) the PBS and the WBS.

The CRISTAL system sits alongside the EDMS and allows for the coordination and execution of activities (also called *tasks*) upon parts (see figure 5). CRISTAL manages the execution of the tasks and collects the output of the resultant measurements and tests in a database. CRISTAL is, in effect, the execution (or enactment) service of instances of the workflows as defined in the WBS, the PBS and the so-called Production Specification parts of the EDMS. As a consequence of this, CRISTAL will need to be able to access the EDMS for Part Definitions and tolerances and for the specification of detector assembly sequences. All the engineering drawings, blueprints, construction procedures, Part Definitions and Part nominal values together with the Workflow Definitions and their compositions will be stored in the EDMS. CRISTAL consequently integrates a production data management and workflow management facility and controls and tracks parts through the manufacturing life cycle from an EDMS-resident design to the final construction and assembly of the detector. It is the CRISTAL software that enables the collection of the physics data necessary for detector construction in a timely and efficient fashion.

In the same way that the PBS has been modelled in the CRISTAL data model (as shown in figure 4) the Workflow Definitions (WBS) of CRISTAL can also be modelled (see figure 6). Figure 6 shows that Workflow Definitions can be either for Composite Activities or Elementary Activities, that Composite activities are made up from Elementary activities or other Composite Activities and that their compositions are shown through ActivityCompositionMembers. Figure 7 shows another subset of the complete CRISTAL data model which concentrates on aspects of integration and association between Workflow and Part Definitions rather than on their individual details. In particular, it outlines the association and execution of a given workflow on a named part in the CRISTAL production and assembly centres. Each Part Definition has at most one Workflow Definition (or assembly sequence) assigned to it and workflows can be reused for several Part Definitions. For example the process for crystal characterisation is very likely to be the same for every type of crystal used in CMS; crystal Workflow Definitions are reused across crystal types.

Therefore a Part Definition is an item in the PBS and the Workflow Definition is the corresponding process definition in the WBS which must be executed when a Part (with that Part Definition) is registered in the CRISTAL system. These definition objects are often referred to as *meta-objects* in modern computing parlance. In practice when a part is created in a centre a workflow object (process) is attached to it, this object being derived from the workflow meta-object.

The CRISTAL data model has been designed so that each assignment of a Workflow Definition to a Part Definition must be for a specific purpose. That purpose could be for the management of documents, for the purposes of CAD/CAM design tracking or, as in this case, for the management of detector assembly. Each purpose has associated with it some Production Conditions (see figure 7). In CMS detector assembly, the Production Conditions are the part objects, commands and locations required as prerequisites for the initiation of a workflow on a part (with a given Part Definition) which results in the capture of Characteristics. Start conditions are essentially the parts required for the workflow to initiate and end conditions the criteria for an activity to complete successfully. The locations (or Applicable Centres) are used to support the distribution of workflow activities. Commands Definitions are described in the following section of this paper.

This method of integrating EDMS and workflow through the definition of a common data model allows other links to be made between the workflow model and the product data model. For example, links can be made to allow version management of different product data models (as proposed by [Ram96]) or to the assignment of workflows to parts for tracking part maintenance. It is this linkage between the part-based world of the PBS and the workflow-based world of the WBS which is novel in the CRISTAL project and which provides the basis for traceability between resources and tasks in the context of CMS assembly and construction.

# 5. Packages and Execution Specifications in CRISTAL

Further aspects of the CRISTAL data model are introduced in this section including the concept of Agents as the mechanisms by which defined workflows are executed and the introduction of execution specifications to include conditions, goals and outcomes on the execution of activities by Agents.

All of the major aspects of CRISTAL functionality (such as Part Definition (PBS), activity definition (WBS), production specification, data formats, execution specification, agent specification and versioning) are specified in the form of packages (see figure 8). Packages are a convenient mechanism to partition the functionality of CRISTAL and they deal with logical subsets of the complete CRISTAL model which are themselves collections of object classes. There are dependencies between the packages, such as between the PBS, the WBS and production specification (as described earlier), and these dependencies provide the required linkage between the classes within different packages. Earlier diagrams have shown this linkage between the product data (PBS) and Workflow Definitions (WBS) and the previous section discussed the importance of maintaining this linkage during CMS construction. This section describes the AgentWorld, ExecutionSpecification and DataFormat packages and the next section that of VersionManagement.

Figure 9 shows the AgentWorld package in CRISTAL. Here Agents are defined as the executors of the CRISTAL workflows, that is, as Humans(Operators or LocalSupervisors), as code supplied by users of CRISTAL or as Instrument definitions. Saved instruction definitions, command definitions and argument definitions can be stored for UserCode or Instrument Agents to define the actual processes involved in execution. As a result of the execution of these instructions, execution results are gathered of a particular DataFormat definition whose detail is shown in the DataFormat package diagram of figure 10. Data formats can be defined in terms of records, Ntuples or fields.

The ExecutionSpecification package (of figure 11) shows the overall assignment of an AgentDefinition to a WorkflowDefinition. This assignment is subject to a set of AgentConditions which define the role of the agent with respect to a Workflow Definition. AgentConditions are made up from Outcome and Goal Definitions which together define the operation of the Workflow by the Agent. Each agent-workflow assignment is established to reach a defined goal and following the execution of its activities, the result is stored as an outcome definition. If the Agent is an instrument then the goal is the command to be completed by the instrument. The next section highlights the importance of versioning in production management and how it is catered for in the CRISTAL data model and covers aspects of the VersionManagement package.

# 6. Handling Versions of Production

The specifications of industrial manufacturing production lines are required to evolve over time. The processes (or workflow activities) which need to be executed to constitute the production line may change in order or in specification. New activities may be inserted into the production line or existing activities may be amended or deleted. In addition to this new products may be specified which may follow an existing production line or may require a completely new specification of the production line.

The production line in CRISTAL (the detector production process) is the set of workflow activities which has to be executed on all the parts constituting the detector for construction. The overall production coordinator monitors the detector production process and, if necessary, amends its definition via the EDMS. Such actions will inevitably result in reallocating tasks and parts and will require new versions of the detector production process. CRISTAL and the EDMS must therefore be able to cope with multiple versions of Part Definitions, task assignments and production processes and allow seamless navigation between these parts and processes. The strategy used in CRISTAL is based on version management of the objects as defined in the EDMS to enable the automatic computation of the amended workflows and the consequent allocation of parts to workflows.

The detector production process is completely described in the current version of the so-called Detector Production Scheme (see figure 12). Amendments to be applied to the Detector Production Process are described in the new version of the Detector Production Scheme. When the amendments take effect, the new version of the Detector Production Scheme becomes the current version and the current version is moved to the version history. At any time there can be at most one new and current version of the Detector Production Scheme. Online, the Detector Production Process is not versioned, it is amended to correspond to the new version of the Detector Production Scheme. Any version in the version history can be browsed. These mechanisms ensure that any changes to the specification of the production processes can be folded through to the active detector production.

# 7. CRISTAL and CMS Assembly

Having discussed the philosophy behind handling versions of the production scheme in a single named CRISTAL system, the overall coordination of CMS assembly using CRISTAL is now explained. CMS assembly will be carried out by groups with responsibilities for sub-detector construction (or sub-sub-detector construction) such as the ECAL, Tracker, HCAL, ECAL-Barrel, ECAL-EndCaps. Each group needs to be only loosely coupled to others for final detector integration and must preserve their autonomy during the assembly process. Therefore, at any point in the overall assembly of CMS there will be several CRISTAL systems active; one for each logical assembly process eg ECAL, ECAL-Barrel and ECAL-EndCaps and these independent CRISTAL systems will be only loosely coupled. Each CRISTAL system will be used to coordinate production and assembly and will have its own PBS graph.

 The complete CMS assembly will be constituted from an hierarchy of these sub-detector CRISTAL systems and will be consistent with the overall CMS PBS tree in EDMS. Each system will be subject to its own versioning strategy following the philosophy outlined above and each system will have a root or point of overall coordination. These roots provide the entry point of one CRISTAL system in the overall CMS assembly PBS tree. The coupling between CRISTAL systems becomes important when a lower level CRISTAL system (eg ECAL-EndCap) is subsumed into a higher level CRISTAL system (eg ECAL) for the purpose of detector integration. Then Part and Workflow Definitions from the lower system can be used in the higher level system; for example, definitions from the ECAL-EndCap CRISTAL system can be used by the ECAL CRISTAL system. This organisation facilitates the complete specification of the assembly of CMS after or while the different sub-detectors are being assembled. Figure 13 diagrammatically represents the hierarchical organisation of CRISTAL systems for CMS.

# 8. Conclusions and project status

Cadim [Cad97] is the commercial product selected by the EDMS task force [Ced97, Ham96] at CERN to support the EDMS functionalities required for the LHC accelerator and experiments. All the EDMS functionalities required by CRISTAL will be put in Cadim. The EDMS will be used in CRISTAL to store the full specification for detector production. It will contain all the definitions of individual parts and will describe what is to be built and how it is to be built. The EDMS specifications will be defined and held centrally. The workflow specification will be derived from the EDMS specification (PBS/ABS/WBS). It will define the process of part characterisation and the complete assembly process from single parts to full-scale detector. In essence, the EDMS will be used as the schema of the workflow meta-objects.

The CRISTAL project has already shown the viability and importance of adopting a dynamic object-oriented approach to the development of complex system software in a rapidly changing application environment and where many implementation choices need to be deferred. Since the project's inception a considerable amount of interest has been generated in meta-models and meta-object description languages [LV97]. Work is in progress within the OMG on the Meta Object Facility (MOF) [OMG96a] that is expected to manage all kinds of meta-models which are relevant to the OMG Architecture. Two which are of particular significance to the CRISTAL project are the Manufacturing's Product Data Management Enablers [OMG96b] and Work Flow Facility[SMB96, OMG97] meta-models. This meta-modelling approach will facilitate further integration between product data management and workflow management thereby providing consistency between design and production and speeding up the process of implementing design changes in a production system.

CMS has selected CRISTAL as the data and process management tool for the construction of the experiment. CRISTAL development was initiated in early 1996 at CERN and a prototype capture tool was developed to allow existing aspects of CMS construction to be incorporated in an object database. The second phase of prototyping and technology evaluation was initiated in the summer of 1996 and the first release of the software will be tested for CMS ECAL in May 98 based on Cadim, CORBA and the Objectivity database [Obj97].

## Acknowledgments

The authors take this opportunity to acknowledge the support of their home institutes. In particular, the support of P Lecoq, J-L Faure and M. Pimia is greatly appreciated. The help of D. Rousset, E. Leonardi and W. Harris in creating the early CRISTAL prototypes is also recognised.

**Figures**

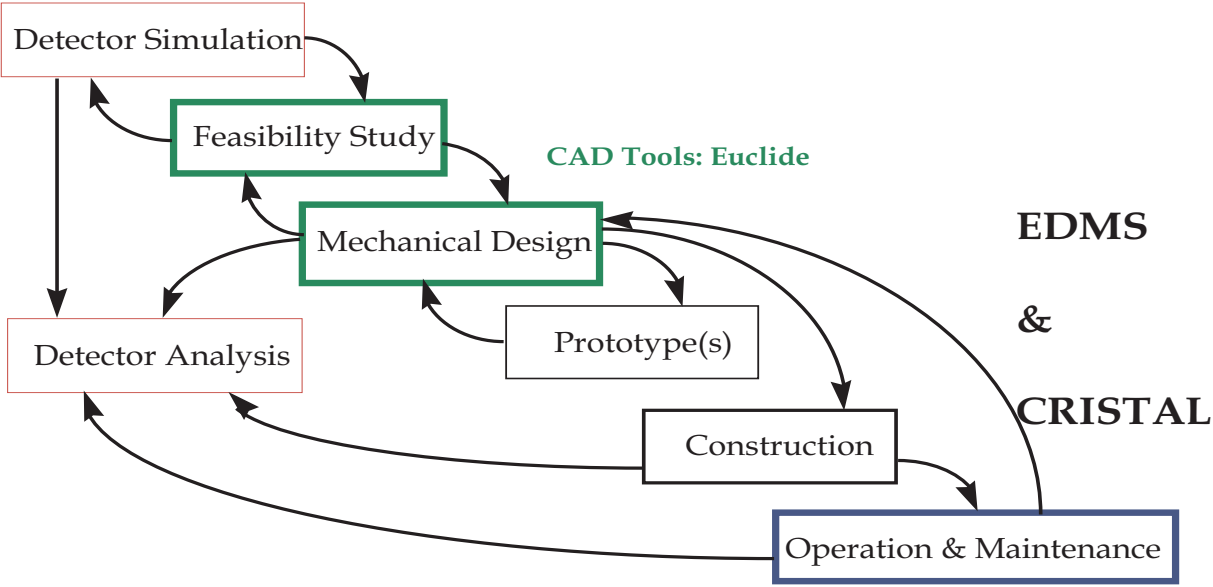

Figure 1: A typical product development lifecycle showing the role of CAD/CAM tools an EDMS and CRISTAL.

| Part Definition (as designed) | Part (as built) |
|---|---|
| Name<br>Sub-Name<br>Type | Part ID<br>PartLocation<br>CurrentActivity |

**n parts produced with this definition** →

| • Characteristic Specifications<br>• Nominal values<br>• Data format Specifications<br>• Instrument Specifications<br>• Activity Specifications | • Physical Characteristics measured by instruments<br>• Comparison with nominal values<br>• Instruments<br>• Activity history |

Figure 2: Separation between Items and Item Definitions.

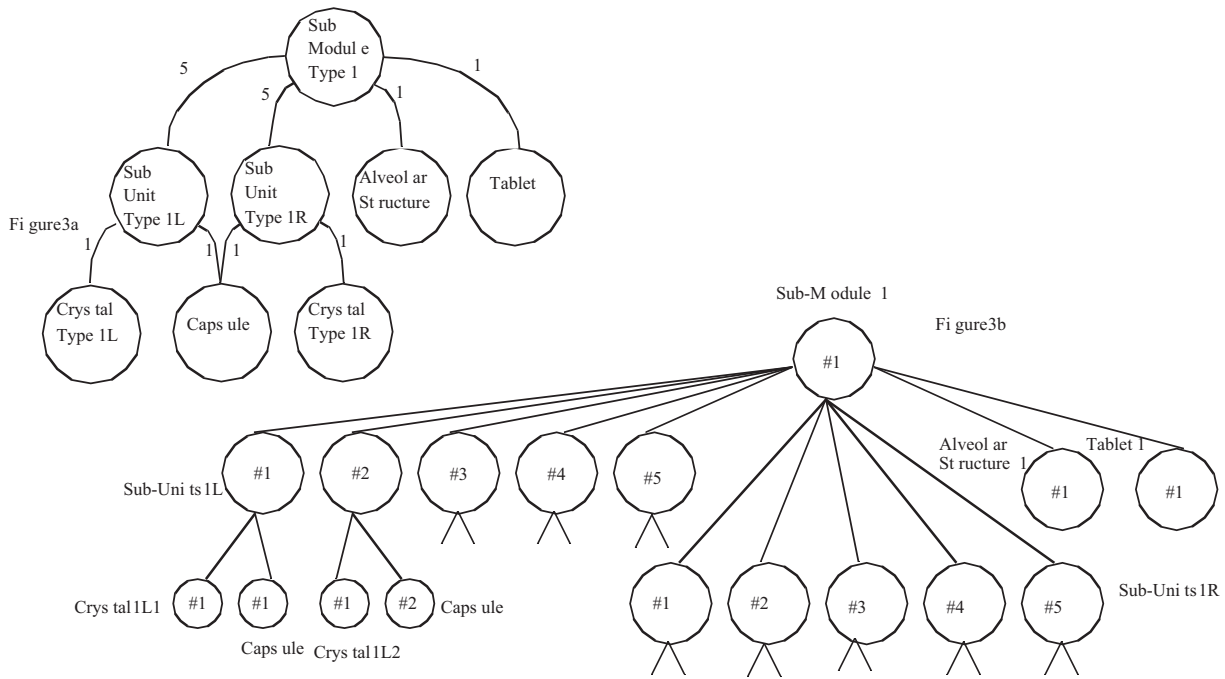

Figure 3 a: The 'As-Designed' PBS graph  & Figure 3b: The 'As-Built' ABS tree for a CMS ECAL Sub-Module.

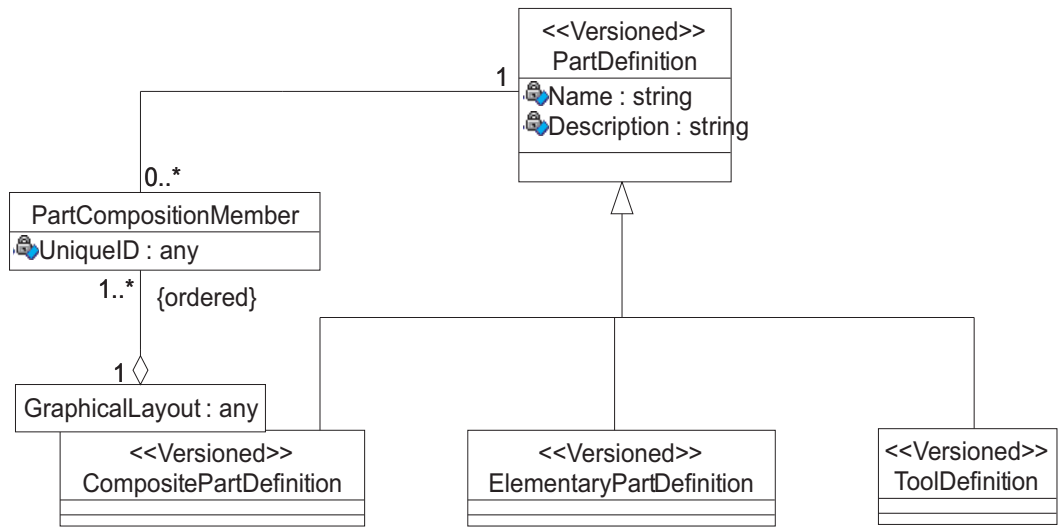

Figure 4: An UML model showing the graph organisation of the Part Definitions and Tool Definition which constitutes the Product Breakdown Structure (PBS).

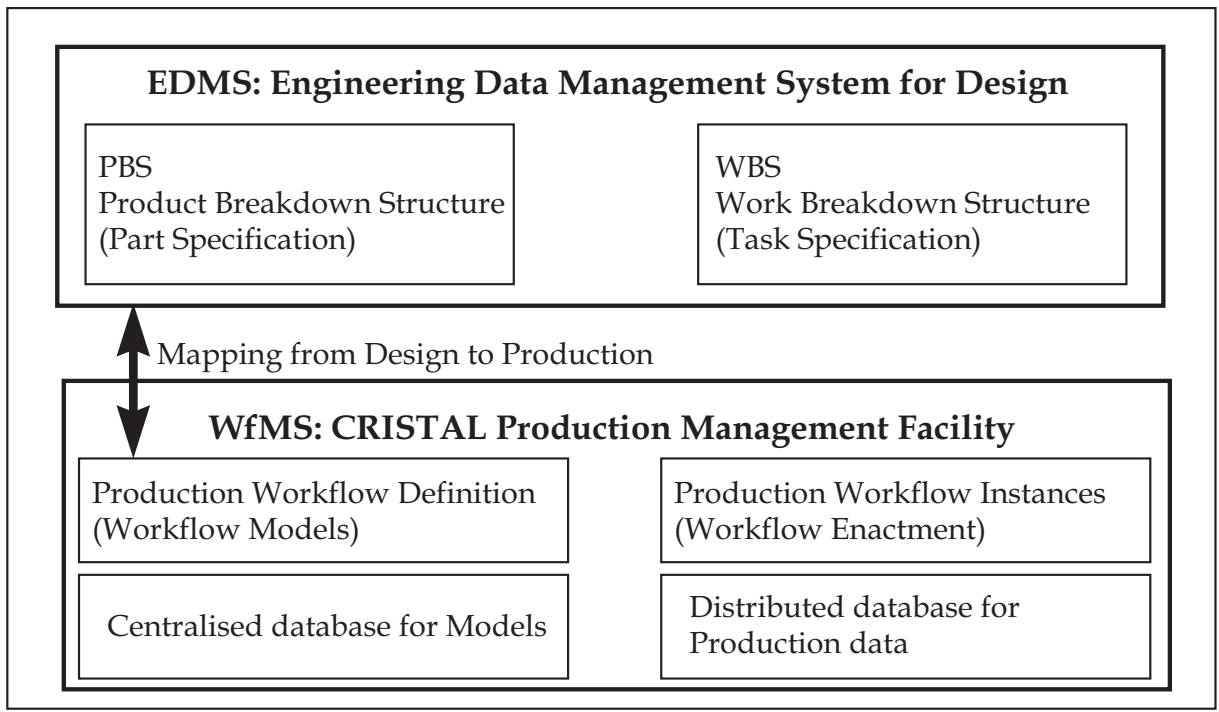

Figure 5: The EDMS constitutes a centrally-available product description and the WfMS constitute the actual processes required to produce the final product(s).

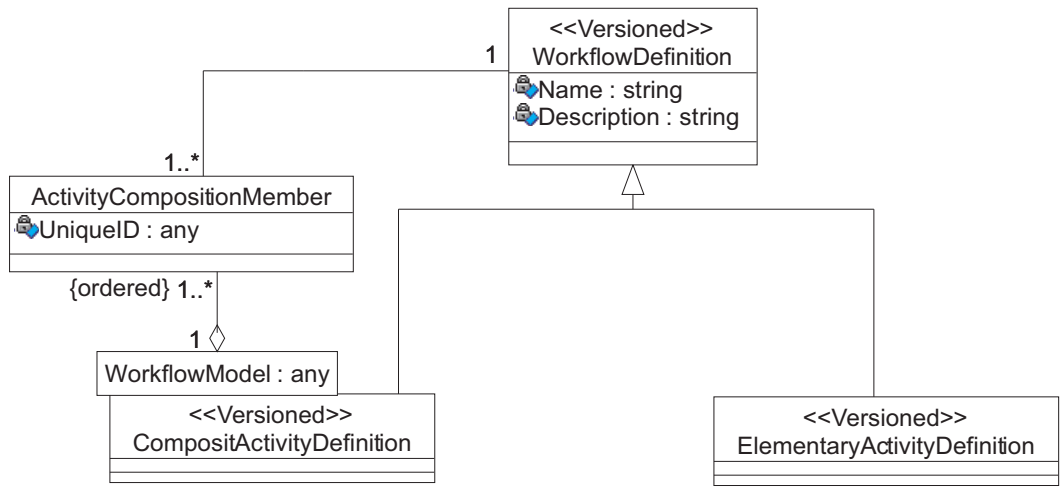

Figure 6: An UML model showing the graph organisation of the Workflow Definitions which constitute the Work Breakdown Structure (WBS).

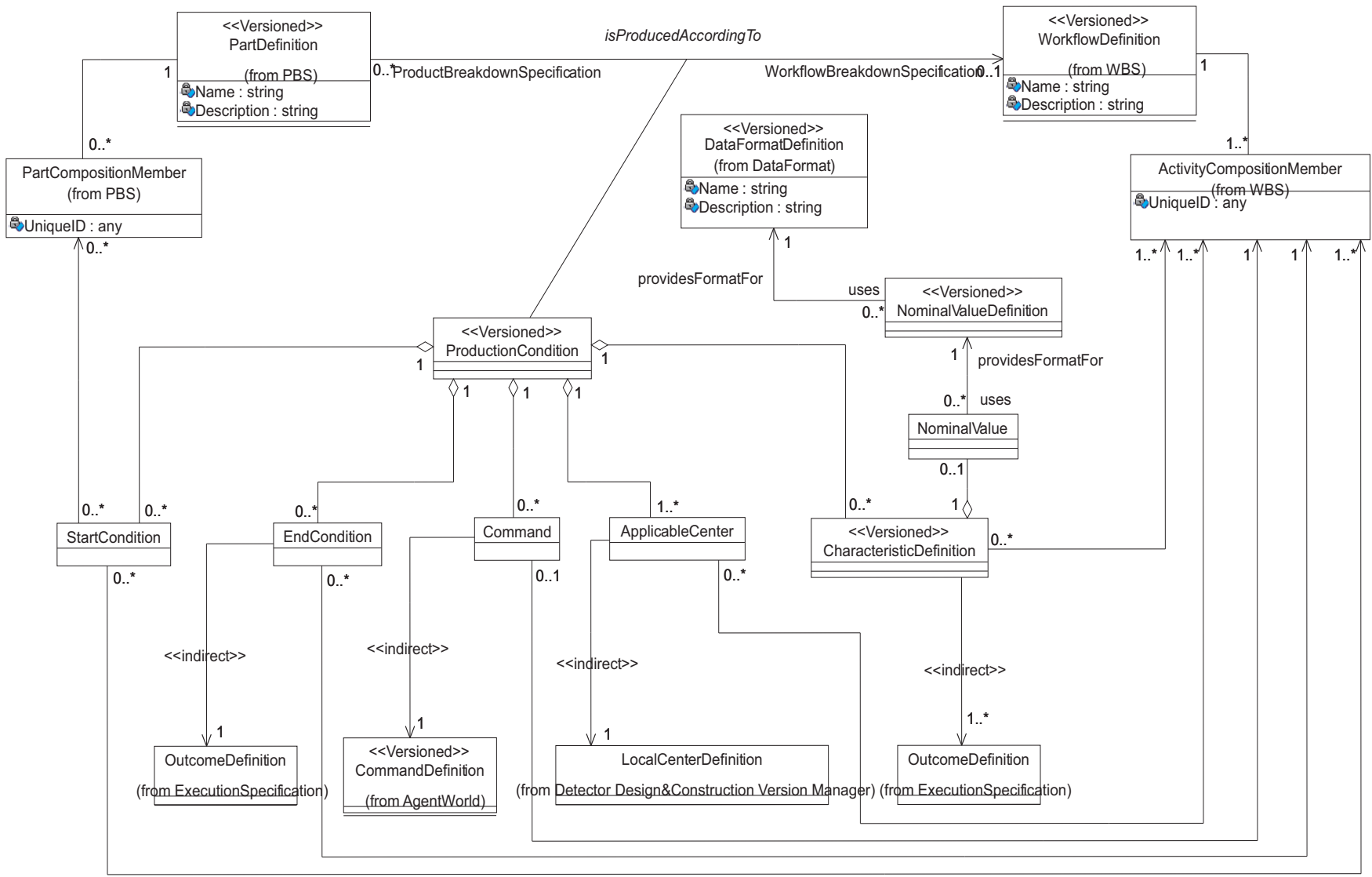

Figure 7: The EDMS data model for CMS detectors design and construction.

Top Level Packages

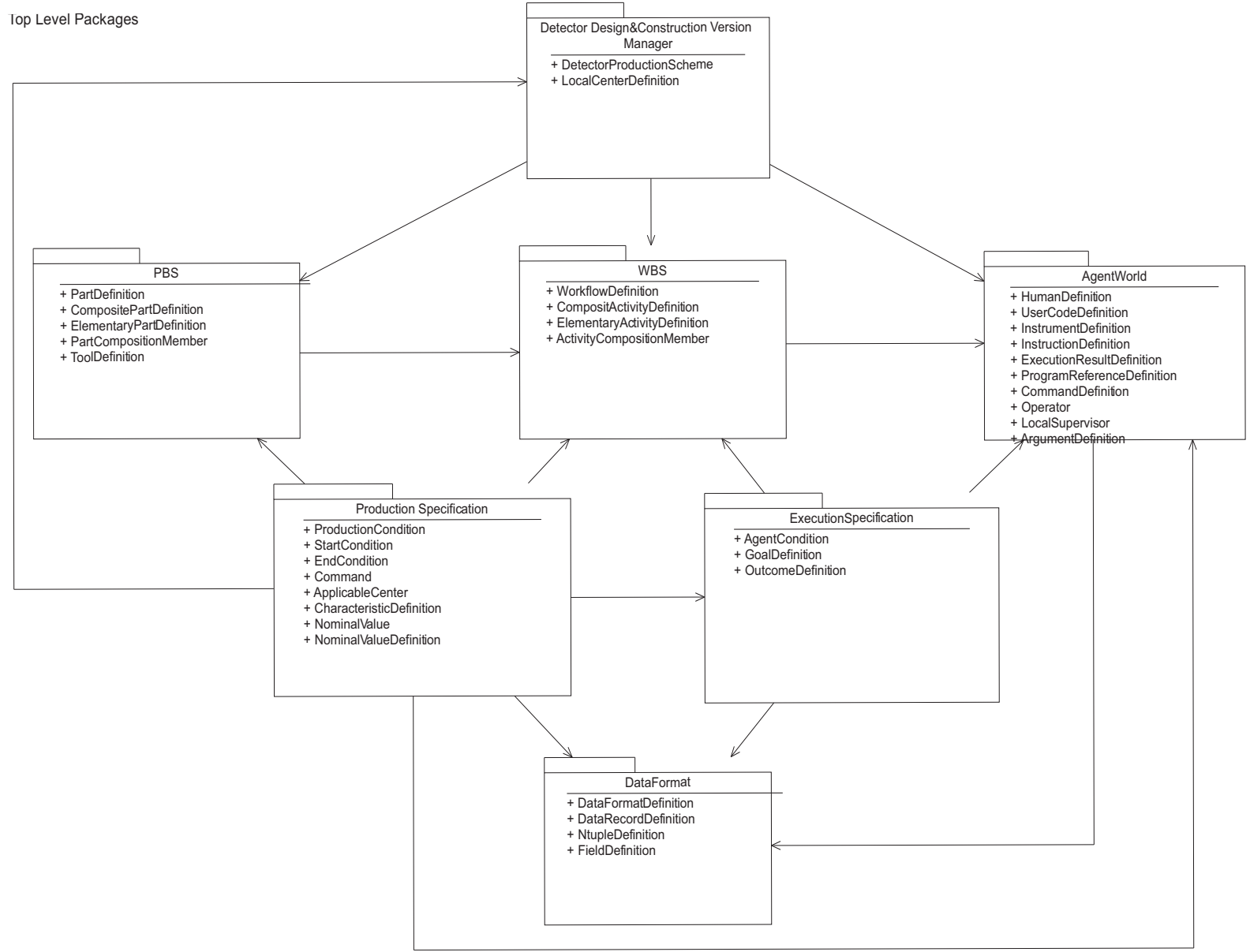

Figure 8: CRISTAL Packages overview diagram.

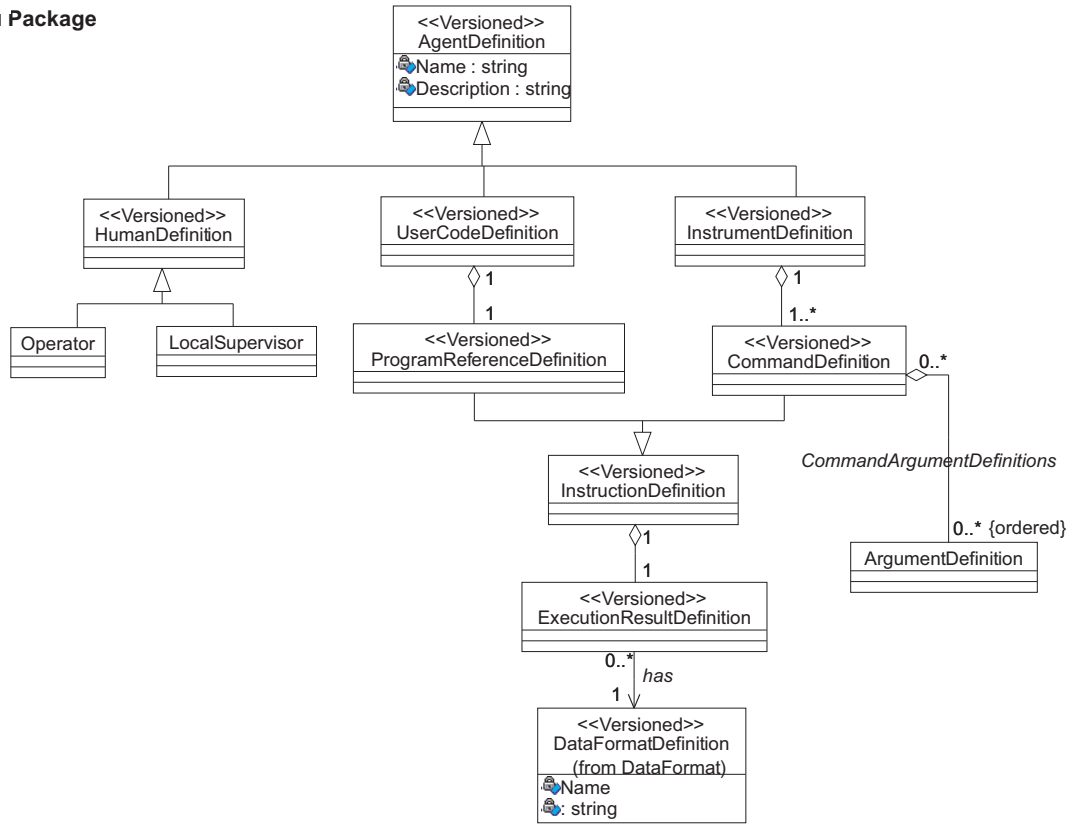

Figure 9: The CRISTAL AgentWorld Package.

## Data Format Package

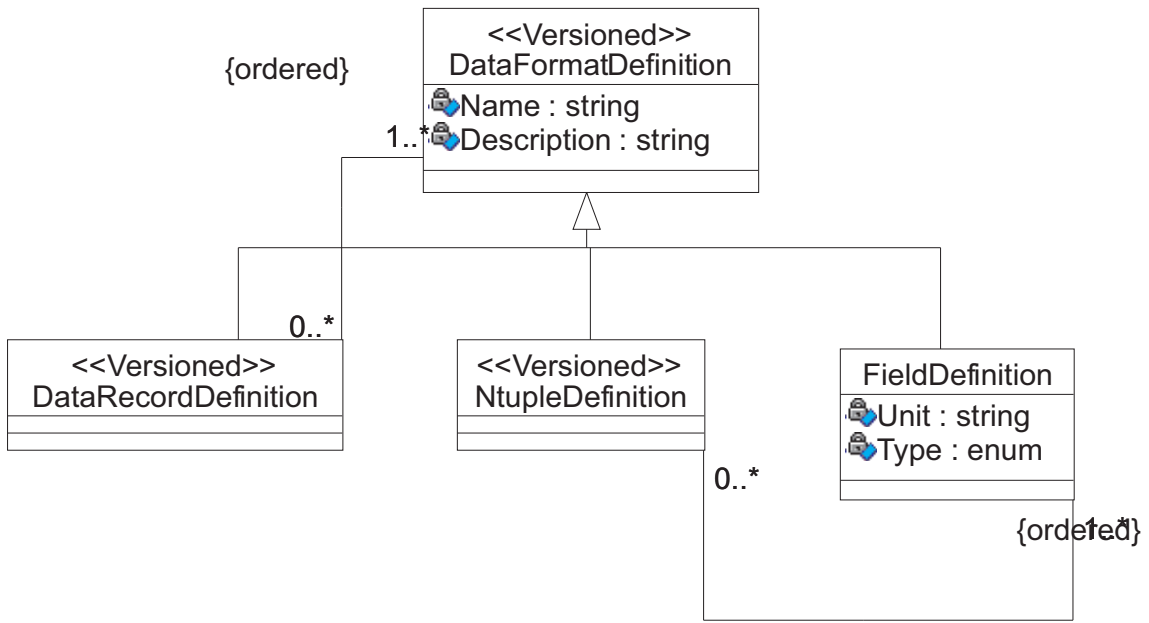

Figure 10: The CRISTAL DataFormats Package.

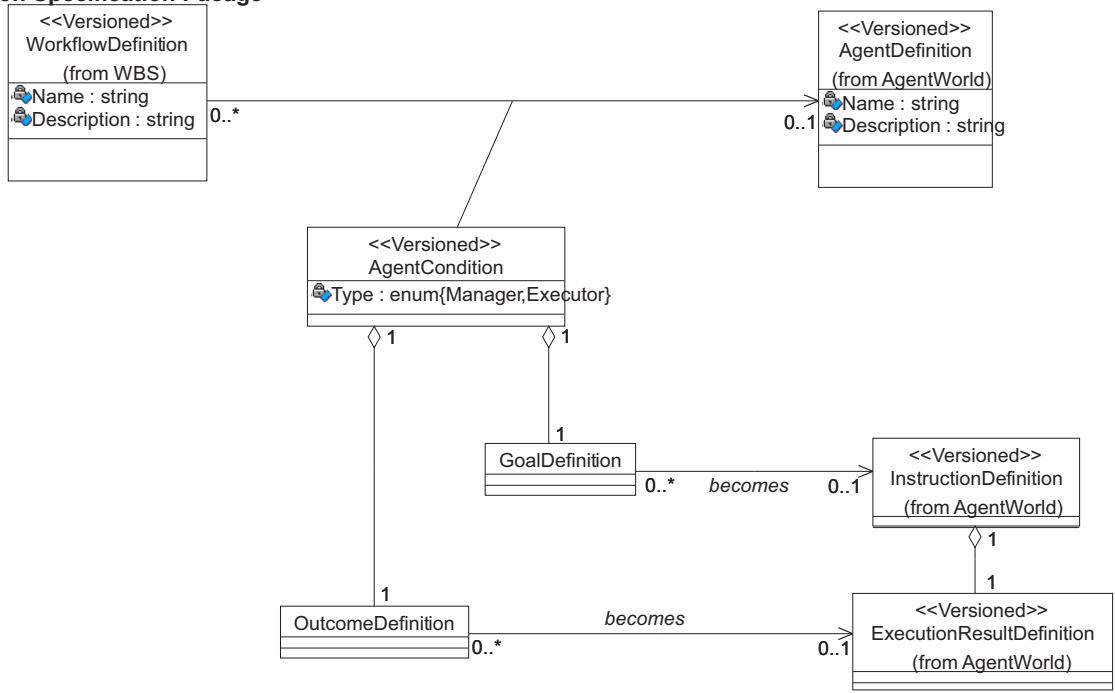

Figure 11: The CRISTAL Execution Specification Package.

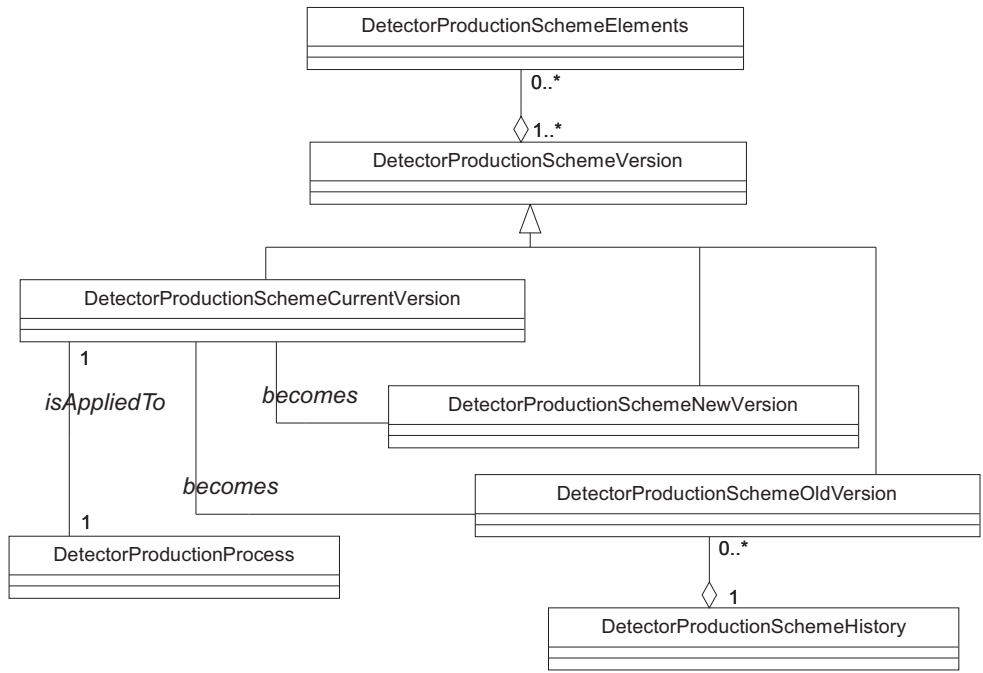

Figure 12: Versioning strategy for the Detector Production Scheme.

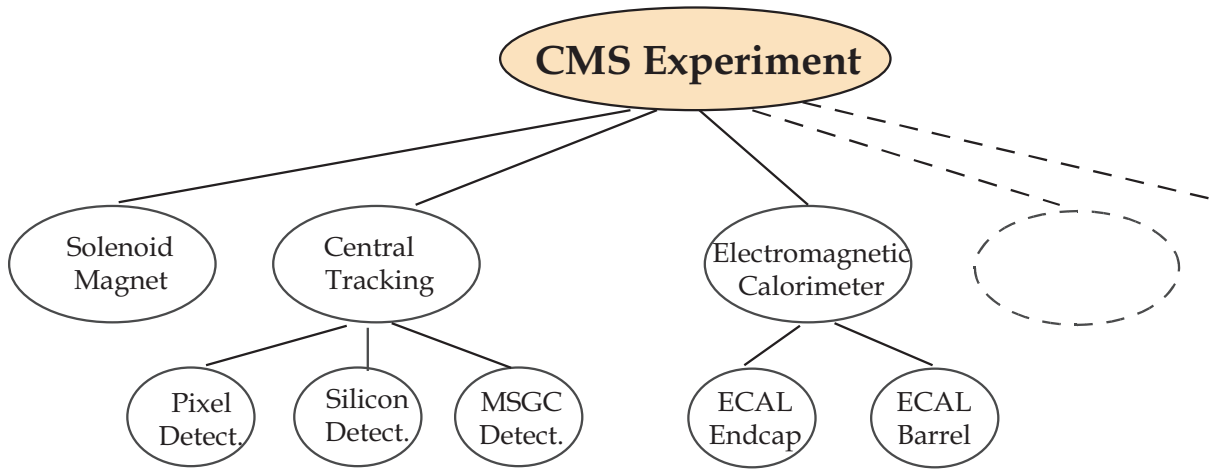

Figure 13: Each node in the tree is a possible CRISTAL system which can access the PBS graph of all its sub-nodes.